\documentclass[10pt, a4]{article}

\usepackage{graphicx, hyperref}
\usepackage{amsfonts, amsmath,amssymb, mathrsfs, wasysym}
\usepackage[english]{babel}
\usepackage[normalem]{ulem}
\usepackage{color,slashed}

\ifx\pdfoutput\undefined
\usepackage[dvips,bookmarks]{hyperref}	
\else
\usepackage{hyperref}	
\fi
\hypersetup{colorlinks,bookmarksopen,bookmarksnumbered,citecolor=rossos,
linkcolor=rossos,pdfstartview=FitH,urlcolor=rossos}

\def\circa#1{\,\raise.3ex\hbox{$#1$\kern-.75em\lower1ex\hbox{$\sim$}}\,}

\numberwithin{equation}{section} 
\usepackage[table]{xcolor}

\definecolor{grigino}{cmyk}{0,0,0,0.2}
\definecolor{mentuccia}{cmyk}{0.4,0,0.3,0.1}
\definecolor{arancino}{cmyk}{0,0.1,0.4,0}
\definecolor{menta}{cmyk}{0.7,0,0.5,0.3}
\definecolor{grigios}{cmyk}{0,0,0,0.5}
\definecolor{bianco}{cmyk}{0,0,0,0}
\definecolor{arancio}{cmyk}{0,0.2,0.6,0}
\definecolor{grigio}{cmyk}{0,0,0,0.1}
\definecolor{rosa}{cmyk}{0,0.1,0.1,0.02}
\definecolor{rosino}{cmyk}{0,0.05,0.05,0.02}
\definecolor{rosas}{cmyk}{0,0.3,0.25,0.05}
\definecolor{celeste}{cmyk}{0.1,0,0,0.02}
\definecolor{giallino}{cmyk}{0,0,0.4,0.02}
\definecolor{rosso}{cmyk}{0,1,1,0.4}
\definecolor{rossos}{cmyk}{0,1,1,0.55}
\definecolor{rossoc}{cmyk}{0,1,1,0.2}
\definecolor{blu}{cmyk}{1,1,0,0.3}
\definecolor{blus}{cmyk}{1,1,0,0.5}
\definecolor{bluc}{cmyk}{1,1,0,0.1}
\definecolor{blucc}{cmyk}{0.7,0.5,0,0}
\definecolor{viola0}{cmyk}{0,0.4,0,0.04}
\definecolor{viola}{cmyk}{0,0.5,0,0.05}
\definecolor{viola2}{cmyk}{0,1,0.2,0.6}
\definecolor{verde}{cmyk}{0.92,0,0.59,0.25}
\definecolor{verdec}{cmyk}{0.92,0,0.59,0.15}
\definecolor{verdecc}{cmyk}{0.42,0,0.8,0.05}
\definecolor{verdes}{cmyk}{0.92,0,0.59,0.4}
\definecolor{verdino}{cmyk}{0.12,0,0.3,0.02}
\definecolor{giallo}{cmyk}{0,0,1,0}
\definecolor{gialloverde}{cmyk}{0.44,0,0.74,0}

\newcommand{\be}{\begin{equation}}
\newcommand{\ee}{\end{equation}}
\newcommand{\bea}{\begin{eqnarray}}
\newcommand{\eea}{\end{eqnarray}}

\newcommand{\newc}{\newcommand}

\newc{\gsim}{\lower.7ex\hbox{$\;\stackrel{\textstyle>}{\sim}\;$}}
\newc{\lsim}{\lower.7ex\hbox{$\;\stackrel{\textstyle<}{\sim}\;$}}

\numberwithin{equation}{section}

\def\fun#1#2{\lower3.6pt\vbox{\baselineskip0pt\lineskip.9pt
  \ialign{$\mathsurround=0pt#1\hfil##\hfil$\crcr#2\crcr\sim\crcr}}}
\def\simgt{\mathrel{\lower0.6ex\hbox{$\buildrel {\textstyle >}
 \over {\scriptstyle \sim}$}}}
\def\simlt{\mathrel{\lower0.6ex\hbox{$\buildrel {\textstyle <}
 \over {\scriptstyle \sim}$}}}

\def\bea{\begin{eqnarray}}
\def\eea{\end{eqnarray}}
\def\be{\begin{equation}}
\def\ee{\end{equation}}
\def\tr{{\rm tr}\,}

\input epsf

\def\be{\begin{equation}}
\def\ee{\end{equation}}
\def\ba{\begin{eqnarray}}
\def\ea{\end{eqnarray}}

\def\tr{{\rm tr}}

\begin{document}

\title{
\vspace{-2.0cm}
\vspace{2.0cm}
{\bf \Large Conformal couplings of  Galileons to other degrees of freedom
}
 \\[2mm]
}
\author{
 Gianmassimo Tasinato
\\[6mm]
\normalsize\it
Institute of Cosmology \& Gravitation, University of Portsmouth,\\
\normalsize\it Dennis Sciama Building, Portsmouth, PO1 3FX, United Kingdom
}

\maketitle

\thispagestyle{empty}

\begin{abstract}
We discuss a  formulation of Galileon actions in terms of matrix determinants in four dimensions. This approach  allows one to straightforwardly determine derivative couplings between Galileons and scalar or vector degrees of freedom that lead to equations of motion with at most two space-time derivatives. We  use this method to  easily 
build generalizations of Galileon set-ups preserving conformal symmetry, finding explicit examples of couplings between Galileons and additional degrees of freedom that preserve  the Galileon conformal invariance. We discuss various physical applications of our 
 method and of our 
results. 
\end{abstract}








\section{Introduction}

Explaining the current  acceleration of our universe is an open theoretical problem. Observational
data are consistent with   cosmological acceleration driven by a cosmological constant, whose
size is however much smaller than what is
  expected
  by an effective field theory approach:  see
\cite{Burgess:2013ara}
  for a review. 
 A possibility is that the cosmological constant  is set to zero by some yet unknown mechanism, while 
 cosmological 
 acceleration is driven by the dynamics of additional degrees of freedom, or
 by a modification of gravity at sufficiently large distances, as reviewed in   \cite{Clifton:2011jh}.
  It is however hard 
 to find consistent theories  that lead to cosmological acceleration, and that at the same time are 
  able to avoid solar system 
 constraints on deviations from  general relativity. 
 
 \smallskip
  
 Recently, a new scenario has been pushed forward,
  based on a scalar Lagrangian with derivative self-interactions respecting a
  Galilean symmetry \cite{Nicolis:2008in}. 
  Although the Galileon Lagrangian contains higher order derivative self-interactions,
   it is designed in such a way that the  corresponding  equations of motion have at most two space-time derivatives, 
   hence avoiding Ostrogradsky instabilities.  These
   field 
    equations have been shown in  \cite{Nicolis:2008in}
    to  admit de Sitter solutions at small scales, that can describe present day acceleration
    even in the absence of scalar potential. Although the scalar field is massless, it does
    not necessarily lead to dangerous long range fifth forces. Derivative self-interactions are able to screen its
    effects in proximity of spherically symmetric sources, realizing the so-called
    Vainshtein screening mechanism \cite{Vainshtein:1972sx}.  
 Galileon theories received much attention over the past few years,
 for their phenomenological 
       consequences. But there have also been interesting theoretical developments suggesting possible
       directions for  consistent
        infrared completions of these theories.        
       Galileon interactions have been explicitly shown to arise in appropriate limits of higher
       dimensional brane probe constructions \cite{deRham:2010eu,Hinterbichler:2010xn}, they appear in  
        compactifications of higher dimensional Lovelock theories \cite{VanAcoleyen:2011mj}, 
       and in the decoupling limit of  a four dimensional theory of massive gravity \cite{deRham:2010ik,deRham:2010kj}. 
       Typically, such constructions add new degrees of freedom, (scalar, vector and/or tensor)
         to the short-distance action that respects  the 
       Galileon symmetry. 
       
       \smallskip 

        The original work \cite{Nicolis:2008in} proposes an intriguing  four dimensional infrared
         completion of the single field Galileon scalar Lagrangian,
        based on a  conformally invariant scalar action. At short distances, and for small values
        of the field, the conformal symmetry 
        reduces to   the Galilean symmetry, hence the conformally invariant action
        approaches the Galilean  one in these limits.
         The 4d conformal group SO(4,2) admits the  SO(4,1) de Sitter 
      group as maximal subgroup:    a non-trivial scalar profile can spontaneously break
      the conformal group to the 
       de Sitter group   \cite{Fubini:1976jm}. Hence, after symmetry breaking, the resulting effective
       action describes scalar fluctuations around the de Sitter configuration, and at short distances
       is well approximated by a Galilean invariant action. Such an approach 
 suggests a possible
        explanation for the observed 
        cosmological acceleration in terms of symmetries: our FRW universe is asymptotically  approaching a maximally symmetric 
        de Sitter configuration,
         resulting  from the  spontaneous breaking of the conformal symmetry
         of an IR complete   Galileon set-up.
                
         \smallskip 

 As discussed already in \cite{Nicolis:2008in}, this proposal is difficult to implement when taking into full account  
         the gravitational backreaction
        of the conformally invariant scalar Lagrangian.  The Weyl rescaling symmetry  that has to be imposed on the gravitational action
        coupled to the conformal system allows one to `gauge away' the Galileon scalar, and work with a purely gravitational action that does not
        admit theoretically  interesting de Sitter 
        solutions.  
 One might wonder whether this negative conclusion might be avoided by considering more general conformal Galileon
        systems, that couple the Galileon to other degrees of freedom (scalars and vectors) in such a way to maintain the conformal
        invariance. When coupling to gravity, there is the possibility that while one of the scalars is gauged away, the  action for the remaining degrees 
        of freedom still respects 
        the structure of a 
        conformal Galileon set-up and is  nevertheless able to generate interesting de Sitter configurations. 
        
        \smallskip 

 Besides this motivation, the question of finding
         conformal couplings of the Galileon to other fields is 
         interesting 
         on its own, since it can reveal  new unforeseen  symmetries or structures in the more general Galileon action, that can suggest
          new ideas 
         to address the problem of explaining current cosmological acceleration. Moreover, conformal Galileon actions 
        are known to be
         more amenable of supersymmetrization  \cite{Khoury:2011da} with respect to standard Galileon actions
          \cite{Koehn:2013hk,Farakos:2013fne}.
  Finding these 
 conformal couplings
  is the scope of this work. 
   We implement a novel method based on a four dimensional analysis of the Galileon action.  We make use of  a very compact, 
        determinantal form of the Galileon action, that renders straightforward to extend it to
        set-ups that
         couple the Galileon to scalar and vector degrees of freedom, in a way that maintain Galilean as well as 
         gauge symmetries. Hence, we show how 
         conformally invariant versions of the resulting  
        actions
        can be easily determined, and express them again in a compact form that renders 
        manifest the symmetries of the system.  
 See also \cite{Fairlie:2011md,Curtright:2012gx,Deffayet:2009mn} for   approaches to 
 Galileons using determinants.
 
 In our conclusions, we analyze various possible applications of our results. We 
        discuss  how 
        they can lead to new actions with derivative couplings among several degrees of freedom (scalar and vector)
        that avoid Ostrogradsky instabilities, and that can be relevant for finding new consistent de Sitter solutions or
        new realizations of screening mechanisms. We speculate how
         our approach  
        to express Galileon invariant actions in a determinantal form  can make more explicit 
         new symmetries and
         structures of the set-ups under consideration. We also discuss problems left  open in this work, as for example 
          the issue of 
          consistently
         including   gravity in these generalized conformal Galileon set-ups.

\section{A convenient way to express Galileon Lagrangians}
We use a convenient way to express Galileon Lagrangians in terms of  matrix determinants.
 Besides leading to  compact expressions, this method renders
 more   explicit  the symmetries
 of the set-ups under consideration. 
  We work in
four dimensions, although our approach can be straightforwardly applied to other  dimensions.  We start with some
 technical formulae that we will need in what follows.
Given a squared $4\times 4$ real matrix $M_{\mu}^{\,\,\nu}$, its  determinant is 
\be\label{det1p}
\det{ M_\mu^{\,\,\nu}}\,=\,-\frac{1}{4!}\epsilon_{\alpha_1\dots\alpha_4}\epsilon^{\beta_1\dots\beta_4}\,M_{\beta_1}^{\,\,\alpha_1}\,\dots\,M_{\beta_4}^{\,\,\alpha_4}\,.
\ee
The contractions between Levi-Civita symbols can be managed using the identity
\be
\epsilon_{\alpha_1\dots\alpha_j\,\alpha_{j+1}\dots\alpha_4}\epsilon^{\alpha_1\dots\alpha_j\,\beta_{j+1}\dots\beta_4}
\,=\,-(4-j)!\,j!\,\delta_{\alpha_{j+1}}^{[\beta_{j+1}}\dots\delta_{\alpha_{4}}^{\beta_{4}]}\,.
\ee
where $[\beta_{j+1} \dots \beta_{4}]$ denotes anti-symmetrization. 

The determinant of $\left(\delta_\mu^{\,\,\nu}+ M_\mu^{\,\,\nu}\right)$ can be also expressed  in terms of traces as follows
\bea\label{expdetr}
\det{\left(\delta_\mu^{\,\,\nu}+ M_\mu^{\,\,\nu}\right)}&=&1+\tr{M}+\frac12\left[ \left(\tr{M} \right)^2-\tr\left(M^2 \right)\right]
+\frac16 \left[ \left(\tr{M} \right)^3- 3\,\tr{M}\,\tr\left({M}^2\right) \,+2\,\tr\left(M^3 \right)\right]\nonumber\\
&&+\frac{1}{24}\,\left[ \left(\tr{M} \right)^4- 6\,\tr{(M^2)}\left(\tr{M} \right)^2\,+3\left(\tr\left(M^2 \right)  \right)^2\,
+8\, \tr{M} \,\tr{(M^3)}
 -6\,\tr\left(M^4 \right) \right]\,.
\eea


 \subsection{The Galileon invariant Lagrangian}\label{sec-giact}
 In order to describe the standard single field Galileon set-up in a compact way, the
  basic building block we need is   the following scalar Lagrangian defined in Minkowski space, $g_{\mu\nu}\,=\,\eta_{\mu\nu}$
 \be\label{Galilag}
{\cal S}_{gal}\,=\,\int d^4 x\, \det{(\delta_\mu^{\,\,\nu}+\frac{1}{\Lambda}\,\Pi_{\mu}^{\,\,\nu}\,- \,\partial_{\mu}\pi \partial^{\nu}\pi)}
\,. 
\ee
This Lagrangian describes the dynamics of a single scalar field $\pi(x)$, and $\Pi_{\mu\nu}\,\equiv\,\partial_\mu \partial_\nu \pi$. Indexes
are raised and lowered with the Minkowski metric.  A parameter
 $\Lambda$ of dimension of mass has  been introduced for dimensional reasons.
   The same Lagrangian 
   can also be rewritten more elegantly as
\be\label{Galilag2}
{\cal S}_{gal}\,=\,\int d^4 x\, \det{(\delta_\mu^{\,\,\nu}-\frac{1}{\Lambda^2}\,e^{{\Lambda\,\pi}{}}\,
\partial_\mu \partial^\nu\, e^{-{\Lambda\, \pi}}
)}\,. 
\ee
Remarkably, this action contains all the Galileon interactions, as we are going to discuss now. 
 
\smallskip

From now on, for simplicity, we  choose units for which $\Lambda=1$. 
Using formula (\ref{det1p}), the determinant contained in  (\ref{Galilag}) can be expanded  as a sum of various contributions.
The contributions that contain only $\Pi_{\mu\nu}$ tensors are total derivatives.  
Indeed, one observes that
\bea
\Pi_{\alpha_1 \beta_1}\dots\Pi_{\alpha_k \beta_k}\,\epsilon^{\alpha_1\dots\alpha_k\,\beta_{k+1}\dots\beta_n}\epsilon^{\beta_1\dots\beta_k\,\beta_{k+1}\dots\beta_n}&=&\partial_{\alpha_1}\left(\partial_{\beta_1} \pi\,\Pi_{\alpha_2 \beta_2}\dots\Pi_{\alpha_k \beta_k}\,\epsilon^{\alpha_1\dots\alpha_k\,\beta_{k+1}\dots\beta_n}\epsilon^{\beta_1\dots\beta_k\,\beta_{k+1}\dots\beta_n}  \right)\nonumber\\
&&-\partial_{\beta_1}\pi\,\partial_{\alpha_1}\left(\Pi_{\alpha_2 \beta_2}\dots\Pi_{\alpha_k \beta_k}\,\epsilon^{\alpha_1\dots\alpha_k\,\beta_{k+1}\dots\beta_n}\epsilon^{\beta_1\dots\beta_k\,\beta_{k+1}\dots\beta_n}  \right)
\nonumber\\
\eea
but the second line vanishes due to  the antisymmetric Levi-Civita symbols. 



The contributions that contain powers of the tensor $(\partial_\mu \pi \partial_\nu \pi)$ higher or equal to two vanish. 
Indeed, one  can write such contributions as

\be
\left(\partial_{\alpha_1} \pi \partial_{\beta_1} \pi\right)\,\left(\partial_{\alpha_2} \pi \partial_{\beta_2} \pi \right)\,K^{(3)}_{\alpha_3\beta_3}
\dots K^{(n)}_{\alpha_n\beta_n}\epsilon^{\alpha_1\alpha_2\dots\alpha_n}\epsilon^{\beta_1\beta_2\dots\beta_n}\
\ee
where the $K^{(i)}_{\alpha_i\beta_i}$  can be $\delta_{\alpha_i\beta_i}$, $\Pi_{\alpha_i\beta_i}$, or $\partial_{\alpha_i}
\pi \partial_{\beta_i} \pi$. For our argument, it is sufficient to know that
the $K^{(i)}_{\alpha_i\beta_i}$  
are symmetric tensors in the indexes $\alpha_i$, $\beta_i$. Then we can write

\bea
&&\left(\partial_{\alpha_1} \pi \partial_{\beta_1} \pi\right)\,\left(\partial_{\alpha_2} \pi \partial_{\beta_2} \pi \right)\,K^{(3)}_{\alpha_3\beta_3}
\dots K^{(n)}_{\alpha_n\beta_n}\epsilon^{\alpha_1\alpha_2\dots\alpha_n}\epsilon^{\beta_1\beta_2\dots\beta_n}\nonumber\\
&&\,=
\left(\partial_{\alpha_2} \pi \partial_{\beta_1} \pi\right)\,\left(\partial_{\alpha_1} \pi \partial_{\beta_2} \pi \right)\,K^{(3)}_{\alpha_3\beta_3}
\dots K^{(n)}_{\alpha_n\beta_n}\epsilon^{\alpha_1\alpha_2\dots\alpha_n}\epsilon^{\beta_1\beta_2\dots\beta_n}
\nonumber\\
&&\,=
-\left(\partial_{\alpha_2} \pi \partial_{\beta_1} \pi\right)\,\left(\partial_{\alpha_1} \pi \partial_{\beta_2} \pi \right)\,K^{(3)}_{\alpha_3\beta_3}
\dots K^{(n)}_{\alpha_n\beta_n}\epsilon^{\alpha_2\alpha_1\dots\alpha_n}\epsilon^{\beta_1\beta_2\dots\beta_n}
\nonumber\\
&&\,=-
\left(\partial_{\alpha_1} \pi \partial_{\beta_1} \pi\right)\,\left(\partial_{\alpha_2} \pi \partial_{\beta_2} \pi \right)\,K^{(3)}_{\alpha_3\beta_3}
\dots K^{(n)}_{\alpha_n\beta_n}\epsilon^{\alpha_1\alpha_2\dots\alpha_n}\epsilon^{\beta_1\beta_2\dots\beta_n}
\label{anticomb}
\eea
Being equal to its opposite, this combination vanishes. 

\smallskip

Hence, when using formula  (\ref{det1p}),  the only non-trivial field dependent  contributions from  the expansion of the matrix
determinant in eq. (\ref{Galilag}) 
are the ones that contain one power of $\partial_\mu
\pi \partial_\nu \pi$, and powers of $\Pi_{\mu\nu}$ from zero to three. These contributions lead to equations of motion with at 
most two space-time derivatives, due to the 
 properties of the antisymmetric Levi-Civita symbol.
 Moreover, 
the action  (\ref{Galilag})
 preserves the  Galileon invariance
$\pi(x)\,\to\,\pi(x)+a+b_\mu x^\mu$. By performing such transformation, the determinant becomes
$$ 
\det{\left[\delta_{\mu}^{\,\,\nu}+\Pi_{\mu}^{\,\,\nu}-(\partial_\mu \pi+b_\mu)(\partial^\nu \pi+b^\nu)\right]}\,.
$$
When evaluating it, one finds that terms that depend on powers of $(\partial_\mu \pi+b_\mu)(\partial_\nu \pi+b_\nu)$
higher or equal to two vanish, for the same arguments developed in  eq. (\ref{anticomb}).
 Then, when considering the terms proportional to $(\partial_\mu \pi+b_\mu)(\partial_\nu \pi+b_\nu)$, one finds that all contributions depending on $b_\mu$ are
 total derivatives: this proves that   (\ref{Galilag}) is invariant under Galileon symmetry.
 
 \smallskip
 
Not surprisingly, expanding  the determinant in   (\ref{Galilag})  
 in terms of traces using formula (\ref{expdetr}),  
one finds a scalar Lagrangian  corresponding exactly to the Galileon 
 Lagrangian discussed in  \cite{Nicolis:2008in}, where the derivative interactions  of different orders 
  combine in Galileon invariant combinations
(dubbed ${\cal L}_2\dots {\cal L}_5$ in \cite{Nicolis:2008in})
  with fixed overall coefficients:
  \bea
  S_{gal}&=&\int d^4 x\,\left[1+2\, {\cal L}_2+3\, {\cal L}_3+2\, {\cal L}_4+\frac56\, {\cal L}_5\right]  \,. 
  \eea
  In order to find a general combination of Galileon Lagrangians, in which each of the previous  contributions has arbitrary
  overall  coefficient, 
 it is sufficient to sum four copies of actions  (\ref{Galilag})   (or (\ref{Galilag2})) in the following way
\be\label{Galisum}
\sum_{i=1}^4\,
\int d^4 x\, \det{(\delta_\mu^{\,\,\nu}+c_i\,e^{{\pi}{}}\,
\partial_\mu \partial^\nu\, e^{-{\pi}}
)}
\ee
with arbitrary parameters $c_i$. 
Expanding the determinants and summing the different contributions, one can fix the relative coefficients  among the
Galileon invariant combinations ${\cal L}_i$
by appropriately choosing the $c_i$ and using Newton identities (similar considerations
have already been done in  \cite{Koyama:2011yg}). Notice that a cosmological constant term  appears, that
can be tuned to zero by adding a bare cosmological constant to the initial action.

\smallskip

To end this subsection, 
it is worth  pointing out that the determinantal  Lagrangian contained in  
(\ref{Galilag}) is not the only one that leads to the  Galileon invariant combinations of \cite{Nicolis:2008in}.
By considering a quadratic shift on the scalar, \be\pi=\frac{c_0}{2}\,x^2+\hat{\pi}\,,\label{qpro}\ee 
for some dimensionful parameter $c_0$, 
one obtains an effective action for the field $\hat{\pi}$ of the form 
\be
S\,=\,\int d^4 x\,\det{\left[ \left(1+c_0\right)\delta_\mu^{\,\,\nu}+\hat{\Pi}_{\mu}^{\,\,\nu}\,- \,\partial_{\mu}\hat{\pi}
 \partial^{\nu}\hat{\pi}
 \,- \,c_0\,\partial_{\mu}\hat{\pi}\,
 x^{\nu}\,- c_0\,\,x_\mu\,
 \partial^{\nu}\hat{\pi}
 -c_0^2\,
 x_\mu\,x^\nu
 \right]}\,.
\ee
Expanding the determinant using formula (\ref{det1p}), one can  show that the various contributions
 can be re-expressed as a sum of Galileon invariant combinations
${\cal L}_i$, with coefficients depending on $c_0$.
  Hence,
the Lagrangian for fluctuations around a quadratic scalar profile as (\ref{qpro}) still obeys 
 the  Galileon symmetry. This fact has indeed been used in \cite{Nicolis:2008in} to characterize 
the Lagrangian for fluctuations around
self-accelerating
de Sitter
 configurations,
for which a scalar profile of the form (\ref{qpro}) solves the background equations
of motion.

\subsection{The conformal Galileon}\label{conf-sing}


The Galileon action  (\ref{Galilag}) is  not the unique  one that leads to equations
 of motion with at most two time derivatives. Simple
extensions exist 
with this property, that can allow one to study 
 systems with larger symmetries, and that possibly reduce to the Galileon symmetric case in appropriate limits.

Consider for example the following action in Minkowski space

\be\label{gennew}
{\cal S}\,=\,\int d^4 x\,\det{
\left(
e^{
\gamma_1 \pi}\,\delta_{\mu}^{\,\,\nu}+
c_2\,e^{\gamma_2\,\pi}\,\Pi_{\mu}^{\,\,\nu}+
c_3\,e^{\gamma_3\,\pi}\,\partial_\mu \pi\partial^\nu \pi
+
c_4\,e^{\gamma_4\,\pi}\,\delta_\mu^{\,\,\nu}\, \partial_\rho \pi\partial^\rho \pi
\right)
}
\ee
for arbitrary parameters $c_i$ and $\gamma_i$.  Expanding the determinant inside the integral by means of eq. (\ref{det1p}) we get
   a sum of different contributions. In this case,  contributions depending only on the tensors
$\Pi_{\mu\nu}$ (weighted by appropriate powers of $e^{\pi}$) are not in general total derivatives.
 On the other hand, for the same arguments developed in eq. (\ref{anticomb}), terms that depend on powers of $\partial_\mu \pi\partial_\nu \pi$
 higher or equal to two vanish. It is straigthforward
   to check that the last term weighted by $c_4$ can be added maintaining  the property that the resulting equations
 of motion contain at most two-space time derivatives; this is due  
 to the properties of the antisymmetric
 Levi-Civita symbol.
  In general the resulting system does  not preserve exact  Galilean invariance, but it 
   might do so 
   in appropriate limits. 

  \bigskip

  In this section, we will show
   that  a special case of the previous action (\ref{gennew}) is invariant under conformal transformations, that reduce to Galileon
   transformations in a short-distance, 
   small $\pi$ limit.  
   The conformal group
is characterized by the following symmetry laws:

\begin{itemize}
\item[-] Dilations $D$ controlled by a parameter $\lambda$
\be
\pi(x)\,\to\,\pi(\lambda \,x)+\ln{\lambda}
\ee
\item[-] Infinitesimal special conformal transformations $K_\mu$ controlled by a vector $c_\mu$
\be
\pi(x)\,\to\,\pi\left(x+ (x^2)\,c-2 \,(c\cdot x)\,x \right)-2\,c_\mu x^\mu
\ee
\item[-] Translations $P_\mu$ controlled by a vector $a_\mu$
\be
\pi(x)\to\pi(x+a)
\ee
\item[-] Boosts $M_{\mu\nu}$ controlled by a tensor $\Lambda_{\mu\nu}$
\be
\pi(x)\to\pi(\Lambda\cdot x)\label{boostr}
\ee
\end{itemize}
The symmetry group that characterizes de Sitter space in four dimension, $SO(4,1)$, is
  a subgroup of the above conformal group $SO(3,2)$. Besides the boosts $M_{\mu\nu}$,
 the generators $S_\mu$ of
  de Sitter algebra 
 are a combination of translations and infinitesimal special confomal trasformations: $S_\mu\,=\,P_\mu-\frac{1}{4}\,H^2\,K_\mu$.
 %


\smallskip

The infinitesimal special conformal transformation requires that $\pi(x)\to\pi(x')-2 c_\mu x^\mu$ with
\bea\label{xptr}
x'^{\rho}&=& x^\rho+ (x^2)\,c^\rho-2 \,(c\cdot x)\,x^\rho 
\eea
for an infinitesimal vector $c^\mu$. This implies that at linear order in $c^\mu$ 
\bea\label{coordtrc}
d x'^\mu&=&\left[ \left(1-2 c_\rho x^\rho\right)\,\delta^{\mu}_{\,\,\nu}+2\left( c^\mu x_\nu-x^\mu\,c_\nu\right)\right]\,d x^\nu\,.
\eea
Calling
 \be\label{defA}{\cal A}^{\mu}_{\,\,\nu}\,\equiv\,\left( c^\mu\,x_\nu-x^\mu\,c_\nu\right)\ee
the antisymmetric combination appearing in eq. (\ref{coordtrc}), we find that
  at leading order in $c^\mu$ the derivative of $\pi$ transforms as
\be
\partial_\mu\,\pi(x)\,\to\,e^{-2 \,c\cdot x}\,\partial'_\mu\,\pi(x')+2\,\partial'_\rho\,\pi(x')\,{\cal A}^{\rho}_{\,\,\mu}-2\,c_\mu\,.
\ee
Hence ($\partial'$ denotes derivatives along the coordinates $x'_\mu$)
\bea
\partial_\mu \pi(x) \partial_\nu \pi(x)&\to&
e^{-4 \,c\cdot x}\,
\partial'_\mu \pi(x') \partial'_\nu \pi(x')-2\left(\partial'_\mu \pi\,c_\nu
+\partial'_\nu \pi\,c_\mu
 \right)+2\left( \partial_\mu \pi\,\partial_\lambda \pi\,{\cal A}^{\lambda}_{\,\,\nu}
 +\partial_\nu \pi\,\partial_\lambda \pi\,{\cal A}^{\lambda}_{\,\,\mu}
 \right)\,,\\
 \partial_\mu\partial_\nu\,\pi(x)
 &\to&e^{-4 \,c\cdot x}\, \partial'_\mu\partial'_\nu\,\pi(x')
 -2\left(\partial'_\mu \pi\,c_\nu
+\partial'_\nu \pi\,c_\mu
 \right)+2\,\eta_{\mu\nu}\,c^\rho \,\partial'_\rho \pi
 +2\left( \partial_{\mu}\partial_{\lambda} \pi\,{\cal A}^{\lambda}_{\,\,\nu}
 +\partial_{\nu}\partial_{ \lambda} \pi\,{\cal A}^{\lambda}_{\,\,\mu}
 \right)\,.
\eea
We will be interested in the combination
\bea \label{combC}
{\cal C}_\mu^{\,\,\nu}&\equiv&
e^{-2\pi}\left[ \partial_\mu\partial^\nu\,\pi-
\partial_\mu \pi \partial^\nu \pi
+\frac12\,\delta_{\mu}^{\,\,\nu}\,\partial^\rho \pi\partial_\rho \pi\right]
\\ \nonumber
&=& \frac12\,\left(\partial e^{-\pi}\right)^2 \,\delta_{\mu}^{\,\,\nu}\,-e^{-\pi}\,\partial_\mu \,\partial^\nu\,e^{-\pi}
\eea
Under conformal transformations, it becomes  
\bea\label{ctrans}
{\cal C'}_\mu^{\,\,\nu}(x')&=&{\cal C}_\mu^{\,\,\nu}(x)+e^{-2\pi}\left[\left( \partial_\mu \partial^\lambda\,\pi- \partial_\mu \pi \partial^\lambda\,\pi\right)\,{\cal A}_{\lambda}^{\,\,\nu}+ {\cal A}_{\mu}^{\,\,\sigma} \left( 
\partial_\sigma \partial^\nu\,\pi- \partial_\sigma \pi \partial^\nu\,\pi
\right)
\right]\\
&\equiv&{\cal C}_\mu^{\,\,\nu}(x)+{\cal B}_{\mu}^{\,\,\lambda}\,{\cal A}_{\lambda}^{\,\,\nu}+ {\cal A}_{\mu}^{\,\,\sigma}
{\cal B}_{\sigma}^{\,\,\nu}
\eea
in the second line, the matrix ${\cal A}$ defined
 in eq. (\ref{defA}) 
is antisymmetric, while the matrixes ${\cal B}$ and ${\cal C}$ are symmetric. 

Consider now the determinant \be\det{\left[\delta_\mu^{\,\,\nu}+{\cal C}_\mu^{\,\,\nu}\right]}\,.\label{detc}\ee
Using eq. (\ref{expdetr}), we can expand the determinant in terms of traces
of ${\cal C}$ and its powers, schematically as $\tr{\left[ {\cal C}^n\right]}$. Under an infinitesimal
  special conformal transformation,  at linear order in $c_\mu$
   each one of these
   traces transforms schematically to $\tr{\left[ {\cal C}^n\right]}\,\to\,
\tr{\left[ {\cal C'}^n\right]}
\,=\,
\tr{\left[ {\cal C}^n\right]}+n\,
\tr{\left[ {\cal C}^{n-1} \,{\cal B} \,{\cal A}\right]}
+n \,\tr{\left[ {\cal C}^{n-1} \,{\cal A} \,{\cal B}\right]}$. But the last two contributions vanish, due to the antisymmetry of ${\cal A}$. 
Hence, each trace involving ${\cal C}$ is invariant under infinitesimal special conformal transformations,
$\tr{\left[ {\cal C'}^n\right]}\,=\,
\tr{\left[ {\cal C}^n\right]}$, 
  and the complete determinant (\ref{detc}) as well. 
 It is also straightforward to prove that the same quantity  is also invariant under dilations. 

 %
%

\bigskip

This implies that the action

\be\label{confact}
{\cal S}_{conf}\,=\,\int d^4 x\, e^{4\pi}\, \det{\left[ \delta_\mu^{\,\,\nu}+c_0\,
e^{-2\pi}\left( \partial_\mu\partial^\nu\,\pi-
\partial_\mu \pi \partial^\nu \pi
+\frac12\,\delta_{\mu}^{\,\,\nu}\,\partial^\rho \pi\partial_\rho \pi\right)\right]}
\ee

\noindent
is invariant under conformal transformations, for arbitrary constant
$c_0$.  Indeed, the 
determinant as we have seen is conformally invariant. The overall factor of $e^{4\pi}$ has been included
to compensate for   the transformation properties of the measure $d^4 x$.  

  This action corresponds to the so-called conformally invariant Galileon system. 
  Expanding the determinant using formula  (\ref{expdetr})
and integrating by parts, one finds a sum of combinations of derivative
interactions involving  $\pi$ (with fixed coefficients) that
 are weighted by different powers of $e^{\pi}$, 
 each of them separately conformally invariant.  They reproduce  
the conformal Galileon Lagrangian described in  \cite{Nicolis:2008in},  obtained
 combining curvature  invariants of a conformally flat metric $g_{\mu\nu}\,\equiv\,e^{2\,\pi}\,\eta_{\mu\nu}$. 
 
 An exception is the conformally invariant contribution  weighted by zero powers of $e^{\pi}$ in the expansion of (\ref{confact}), that turns out 
 to be a total derivative: already  in   \cite{Nicolis:2008in} such a term had to be put by  
  hand, or obtained as a suitable limit of a $d\neq4$ combination. Similar considerations can be  done here.

\smallskip

In order to find the most general conformal Lagrangian with arbitrary coefficients in front of  the different conformally
invariant contributions, it
is enough to proceed exactly as in the case of standard Galileon (see the discussion around eq. (\ref{Galisum})):  
 sum four copies of the action (\ref{confact}), with  different arbitrary coefficients $c_0$, and then properly
 use Newton identities to fix the preferred coefficient in front of each 
 conformally invariant
 combination.

 \bigskip



As reviewed in the introduction, 
the main motivation of \cite{Nicolis:2008in} for studying  a conformally invariant 
 Galileon 
set-up
  is 
  to provide
  a consistent IR completion of Galileon actions in four dimensions.  In this view, motivated
 by the results of \cite{Fubini:1976jm},  the fact that our universe is apparently approaching a de Sitter
 configuration   can be 
  interpreted
  as a  process of spontaneous symmetry breaking. Namely,
  the spontaneous breakdown of  the SO(4,2) conformal symmetry to an  SO(4,1) de Sitter
  symmetry 
 by means of a non-trivial profile for 
 the scalar $\pi$ (acting as dilaton). An example of such profile is 
 \be
 e^{\pi(x)}\,=\,\frac{1}{1+\frac{1}{4}\,H^2 x^2}
 \ee
 with $H$ corresponding to the de Sitter scale. 
This scalar profile   preserves the subgroup of de Sitter symmetries listed between eqs. (\ref{boostr}) and (\ref{xptr}). 
 Hence by general grounds one expects that it solves the equations of motion for the conformal symmetric Lagrangian, by
 appropriately choosing the free parameters that characterizes it. The Lagrangian for fluctuations around this solution respects
 the  de Sitter
 symmetry, and in a  short distance/small field limit
 acquires a 
    Galileon symmetry.
   On the other hand, it provides a potentially consistent IR completion of the Galileon
   action valid  also at large distances. There are issues 
   with this proposal
   when
  coupling the action to gravity, as explained in \cite{Nicolis:2008in} and reviewed in the introduction, that can motivate
  the question of determining conformal couplings of the Galileon to additional degrees of freedom, as scalars 
  and vectors. We will address this question in what follows.

\section{The bi-Galileon and its conformal version}

An interesting feature of our method
 is that it allows one  
 to generalize the previous action and 
determine  
couplings between 
Galileons and other fields,
  respecting    Galileon or conformal symmetries.
In this section, we will
 discuss couplings between Galileon and scalar fields, focussing on 
  the case of the bi-Galileon \cite{Padilla:2010de,Deffayet:2010zh}. 

\smallskip

Consider  two scalar fields $\pi$ and $\sigma $: we require that both of them respect Galileon symmetries
\be\label{bigals}
\pi\to\pi+c^{(\pi)}+b^{(\pi)}_\mu x^\mu\,\hskip1cm\,\sigma\to \sigma +c^{(\sigma)}+b^{(\sigma)}_\mu x^\mu\,.
\ee
For the moment, we choose
 arbitrary parameters $b_\mu^{(i)}$, $c^{(i)}$. 
The general structure of an action that couples these scalar fields in a way that preserves
 the   symmetry (\ref{bigals}) 
is the following  
\be\label{bigal1}
{\cal S}_{bi-gal}\,=\,\int d^4 \,x\,\left(\det{\left[ 
\delta_{\mu}^{\,\, \nu}+ e^{\pi} 
\partial_\mu \partial^\nu e^{-\pi}+ \partial_{\mu}\partial^{\nu}\,\sigma
\right]}
+\det{\left[ 
\delta_{\mu}^{\,\, \nu}+ e^{\sigma} 
\partial_\mu \partial^\nu e^{-\sigma}+ \partial_{\mu}\partial^{\nu}\,\pi
\right]}
\right)\,.
\ee
  Using the arguments we developed in the single Galileon case, it is straightforward
to check that this action respects the Galileon invariance (\ref{bigals}). 
 Expanding the determinants, one obtains
  all the interactions that characterize bi-Galileon theories of  \cite{Padilla:2010de}.
    There appear 
 various Galileon-invariant combinations with fixed coefficients: 
 to determine a general action with arbitrary coefficients in front of those
 combinations
one proceeds as explained around eq (\ref{Galisum}), summing various copies of  (\ref{bigal1})
with suitable coefficients weighting  the second and third term inside each of the two determinants.
 Multi-Galileon
theories can  be straigthforwardly expressed in the same manner,  by adding more terms
 in the action depending on the additional fields, with the same structure as the ones  above.
 %

\smallskip

It is not difficult to 
find a conformal version of the bi-Galileon theory,   symmetric under  the conformal symmetry discussed 
 in section \ref{conf-sing},   that applies to both scalars $\pi$ and $\sigma$. It is sufficient to build analogues
 of the combination ${\cal C}_\mu^{\,\,\nu}$ of eq. (\ref{combC}) for 
 each  scalar $\pi$, $\sigma$, and linearly combine them.  
 An example is 
   the following
 action 
 \bea\label{confact2}
 {S}_{conf}&=&\int d^4 x
 \,e^{4\pi}\, \det \Big[ \delta_\mu^{\,\,\nu}+
a_1\,e^{-2\pi}\left( \partial_\mu\partial^\nu\,\pi-
\partial_\mu \pi \partial^\nu \pi
+\frac12\,\delta_{\mu}^{\,\,\nu}\,\partial^\rho \pi\partial_\rho \pi\right)
\nonumber
\\&&\hskip2.6cm+
a_2\,e^{-2\sigma}\left( \partial_\mu\partial^\nu\,\sigma-
\partial_\mu \sigma \partial^\nu \sigma
+\frac12\,\delta_{\mu}^{\,\,\nu}\,\partial^\rho \sigma\partial_\rho \sigma\right)\Big]
 \eea
 for arbitrary coefficients $a_1$, $a_2$. One proves that this action is conformally invariant repeating
  exactly 
  the 
 same steps we already made for the single scalar case, before eq (\ref{confact}).
  indeed, the determinant is by construction conformally invariant.  The overall coefficient
  $e^{4 \pi}$ in front of it have been placed to take care 
  of the transformation of the measure. For simplicity,  for this purpose we have chosen  to put an overall
    factor depending only on $\pi$, 
  but a different choice depending also on $\sigma$ can also be made.   
  Expanding the determinant of (\ref{confact2}), one finds a sum of several conformally
  invariant combinations, each of them weighted by different
  powers of $e^{\pi}$, $e^{\sigma}$. 
   The advantage
 of (\ref{bigals}) is that it condenses in a  compact formula all such combinations.   
  Notice that, in the small
 field and short distance limit, one recovers  a Galileon symmetric set-up in which both fields $\pi$, $\sigma$
 are invariant under a Galileon symmetry with the {\it same} parameters appearing 
 in eq (\ref{bigals}), $b^{(\pi)}_\mu=b^{(\sigma)}_\mu$, 
  $c^{(\pi)}\,=\,c^{(\sigma)}$.

\bigskip

   As for the case of single Galileon field, the conformal symmetry can be spontaneously broken by a non-trivial profile for one (or both)
   the scalars $\pi$, $\sigma$. If the profile respects de Sitter symmetry,
   a subgroup of the full conformal symmetry, one expects 
    to have it as particular solution of the conformal Lagrangian: fluctuations around such profile will respect de Sitter symmetry and,
     in the small-scale limit, will be governed by a bi-Galileon invariant theory expanded around a de Sitter configuration.
     It would be interesting to understand whether the resulting multi-scalar system can be  consistently coupled 
   to dynamical gravity preserving   the conformal Galileon scalar
    symmetry, and at the same time admitting non-trivial and interesting de Sitter
   solutions. This topic is left for future work. 
   
\section{Conformal couplings of Galileons  to vectors}

Another sector to which Galileons can couple are vector degrees of freedom: also this
case can also be discussed by means of an elegant determinantal formulation. Galileons coupling
to vectors are theoretically well 
motivated, since they arise
 in a suitable decoupling limit of massive gravity \cite{Koyama:2011wx,Gabadadze:2013ria,Ondo:2013wka}. 
In order to couple Galileons to vectors, some 
 further formula for conveniently handling
 determinants will be needed.
Consider an abelian vector field $A_\mu$ with field strength $F_{\mu\nu}\,\equiv\,\partial_\mu A_\nu-\partial_\nu A_\mu$ invariant
under an abelian $U(1)$ symmetry $A_\mu\,\to\,A_\mu+\partial_\mu \xi$. The  field strength is an antisymmetric
tensor in the indexes $(\mu,\,\nu)$.  We will be interested in matrixes 
   $M_\mu^{\,\,\nu}$ that
   combine symmetric and antisymmetric parts, of
    the form
\be
M_\mu^{\,\,\nu}\,=\,S_\mu^{\,\,\nu}+F_{\mu}^{\,\,\nu}
\ee
where  $S_{\mu\nu}$ an arbitrary symmetric matrix. 
Writing schematically $\left(\delta_\mu^{\,\,\nu}+M_\mu^{\,\,\nu}\right)\,=\,\left(1+M\right)$, 
we can write
\bea
\det{\left(1+M\right)}&=& \det{\left(1+S\right)}-\frac{1}{2} \tr{F^2}+\frac18\left[ \left(\tr{F^2} \right)^2-2\tr{F^4}\right]
\nonumber\\ \nonumber 
&&+\tr(S F^2)-\frac{1}{2} \tr{F^2}\tr{S}\,\\
&&+\tr{S}\,\tr(S F^2)-\tr(S^2 F^2)-\frac14\,\tr{F^2} \left[ \left(\tr{S} \right)^2-\tr{S^2}\right]-\frac12\tr{(SFSF)}\,.
\label{vecdet}
\eea
By means of this result, it is straightforward to couple scalars $\pi$ to vectors in a way that preserve Galileon symmetries.
To build  a set-up that preserves both the scalar Galilean symmetry 
 $\pi\,\to\,\pi+b_\mu x^\mu+c$ 
 and vector abelian symmetry,  we choose 
  $S_{\mu}^{\,\,\nu}\,=\,\Pi_{\mu}^{\,\,\nu}$, and write the action as
  \be\label{lagvec1}
  {\cal S}_{vec}\,=\,\int d^4 x\,\det{\left(\delta_{\mu}^{\,\,\nu}+ \Pi_{\mu}^{\,\,\nu}+ F_{\mu}^{\,\,\nu}\right)}\,.
  \ee
  Expanding the determinant  with the help of eq (\ref{vecdet}) (using the symmetric matrix $\Pi_{\mu}^{\,\,\nu}$
   in place of $S_{\mu}^{\,\,\nu}$), one finds all the vector Galileon combinations  
  determined in \cite{Deffayet:2010zh,Tasinato:2013oja}  with fixed coefficients. These  combinations couple 
  the vector to the scalar preserving Galileon and gauge symmetries, and
   lead to equations of motion with at most 
  two space-time derivatives. 
   Combining various copies of the action (\ref{lagvec1}) with different choices of the coefficients
  in front of the second and third term inside
  the determinant, one can reproduce the Galileon and gauge symmetric vector-scalar combinations of \cite{Tasinato:2013oja}, with arbitrary
  coefficients in front of each of them.  
   
   \bigskip
   
   We can go beyond the Galileon invariant case, and look for an action coupling scalars to vectors 
   by means of 
    derivative interactions
    breaking the Galileon symmetry, and that  nevertheless maintain equations of motion with at most second space-time derivatives.
    An example of such an action 
    is

\be\label{gennew1}
{\cal S}\,=\,\int d^4 x\,\det{
\left(
e^{
\gamma_1 \pi}\,\delta_{\mu}^{\,\,\nu}+
c_2\,e^{\gamma_2\,\pi}\,\Pi_{\mu}^{\,\,\nu}+
c_3\,e^{\gamma_3\,\pi}\,\partial_\mu \pi\partial^\nu \pi
+
c_4\,e^{\gamma_4\,\pi}\,\delta_\mu^{\,\,\nu}\, \partial_\rho \pi\partial^\rho \pi
+c_5 \,e^{\gamma_5\,\pi}\,F_{\mu\nu}\right)
}
    \ee
    for arbitrary parameters $c_i$, $\gamma_i$. 
    By means of the tools explained in the previous
    sections, this action can be shown to lead to equations of motion with at most two space-time derivatives,
    although it breaks the Galileon symmetry (while maintaing the abelian gauge symmetry on the vectors). 
    
    \bigskip
    
   A particularly interesting special case of (\ref{gennew1}) is an action that preserves
   conformal symmetry.
     Under 
    infinitesimal special
   conformal transformations,
    the field strength $F_{\mu\nu}$ transforms as (see e.g. \cite{Jackiw:2011vz})
   \be
   e^{-2\pi} {F_{\mu\nu}}\to e^{-2\pi} {F_{\mu\nu}}+2\, e^{-2\pi} \,\left({\cal A}_{\mu}^{\,\,\lambda} F_{\lambda \nu}+ 
   {\cal A}_{\nu}^{\,\,\lambda} F_{\lambda \mu}
   \right)
   \ee
   with the antisymmetric matrix ${\cal A}_{\mu}^{\,\,\nu} $ 
    defined in eq. (\ref{defA}). 
   Using this fact, an action that preserves conformal invariance results
   \be\label{lagvec2}
  {\cal S}_{conf}\,=\,
  \int d^4 x\,e^{4\pi}\, \det{\left[ \delta_\mu^{\,\,\nu}+c_1\,
e^{-2\pi}\left( \partial_\mu\partial^\nu\,\pi-
\partial_\mu \pi \partial^\nu \pi
+\frac12\,\delta_{\mu}^{\,\,\nu}\,\partial^\rho \pi\partial_\rho \pi\right) +c_2\, e^{-2\pi }F_{\mu\nu}\right]}
  %
  \ee
  for arbitrary coefficients $c_1$, $c_2$. By means of equations (\ref{expdetr}) and (\ref{vecdet})
  the determinant inside eq (\ref{lagvec2}) can 
   be expanded in terms of traces. As for the case of the scalar Galileon, using the arguments
    developed in the previous sections 
     it is straightforward 
  to prove that each trace is invariant under infinitesimal special conformal transformations 
  and dilations,  hence the entire
  determinant is conformally invariant.
  The conformally invariant action can be expanded in a sum of several combinations weighted
  by different powers of $e^{\pi}$, each of them invariant under conformal and abelian symmetries: eq  (\ref{lagvec2}) 
   condenses in a compact expression these combinations.  The previous conformal vector-scalar Lagrangian
  can also be straightforwardly extended to a multi-scalar version, 
   for example  
  adding inside the determinant 
 a scalar degree of freedom $\sigma$ with  a combination as the ${\cal C}_{\mu}^{\,\nu}$
    of eq. (\ref{combC}) that preserves conformal symmetry by itself.

  \smallskip
  
Hence, also the conformal   version of Galileon couplings to vectors can also be elegantly described in a determinantal
form. 
 It would be interesting to study whether the resulting system   can 
 improve on 
 the instabilities  of de Sitter configurations driven by vector fields in a Galileon set up, analyzed in \cite{Tasinato:2013oja}.
   
\section{Outlook}\label{sec-disc}
In this work we discussed  an elegant formulation of Galileon actions in terms of matrix determinants  in four dimensions.
This formulation  allows one  to  straightforwardly
determine  derivative couplings between Galileons and scalar and vector degrees of freedom
that lead to equations of motion with at most two space-time derivatives.
We have shown how to use this method 
to build generalizations of Galileon set-ups
preserving conformal symmetry, finding the first examples of couplings between Galileons and additional
 degrees of freedom able to maintain the Galileon conformal invariance.  
 
 \smallskip
 
 Having a novel point of view to determine  consistent derivative couplings of scalars to other degrees of freedom 
 is interesting, 
 because it allows one to generalize known results in new directions. 
 Let us discuss some examples of future investigations.
 
  Recently it has been shown that Galileon systems satisfy a duality relation \cite{deRham:2013hsa} (see also 
  \cite{Creminelli:2013fxa}) associated with a field dependent coordinate transformation $x^\mu\,\to\,x^\mu+\partial^\mu\,\pi$. Using this
  fact, it is straightforward to prove one of the 
  results of section \ref{sec-giact}, namely that the action 
  $$
  S\,=\,
  \int d^4 x\,\det{\left(\delta_\mu^{\,\,\nu}+\partial_\mu\partial^\nu\,\pi\right)}
  $$
 is trivial since it leads to non-dynamical equations of motion for the scalar $\pi$. This fact  can be shown 
 by expanding the determinant and showing that the various terms assemble in total derivatives,
 as explained  in section \ref{sec-giact}. Or more simply, observing
  that the determinant inside the previous integral is nothing but the Jacobean of the coordinate transformation above, hence
  it is not surprise that this action  corresponds to a non-dynamical theory. One can extend the analysis studying the effect of the previous
   coordinate transformation
  for  {dynamical} cases using 
  our determinantal approach, studying generalizations of the Galileon duality to the case of new couplings between Galileons
  and 
  other degrees of freedom. We will 
  develop this investigation elsewhere.

  An important
  consequence of derivative self-couplings of scalar fields is the realization of screening mechanisms
  as the Vainsthein effect, 
 in which non-linearities of the scalar action induce strong interactions in proximity of spherically 
 symmetric sources that are able
 to hide the effects of light scalar fields. Our determinantal formulation of actions with derivative
 couplings render the analysis of spherically symmetric set-ups particularly straightforward, and allow one 
 to generalize known results in various ways, 
 that we will present
   in a separate work.

 One main motivation for studying the conformally invariant actions 
considered 
   in our work 
 is the proposal pushed  forward in \cite{Nicolis:2008in}, that we reviewed in the Introduction. Namely, 
 the possibility  
 to
     find   infrared completions 
 of Galileon systems that admit de Sitter configurations able to explain the current acceleration of our universe. 
  In the single scalar Galileon case, as argued already in \cite{Nicolis:2008in},
   such proposal has problems once the dynamics of gravity coupling to the conformal action is  taken  
     into account. The reason is that Weyl symmetry is able to gauge away the scalar,  leaving with a system that does
     not admit  interesting de Sitter configurations. The hope is to improve 
      the situation 
     adding new degrees of freedom
      coupling to the Galileon, that respect the Galilean conformal symmetry.
      When coupling to gravity, there is the possibility that while one of the scalars is gauged away, 
       the action for the remaining degrees of freedom still respects the structure of a conformal Galileon set-up and is able to generate interesting de Sitter configurations.
        Conformal couplings
      of gravity to Galileons are not however  easy to 
       include  in the elegant determinantal formulation adopted in this work. Articles as \cite{Deser:1998rj} offer useful ideas 
       on this respect, that we leave for developments in
         a future work.

\subsection*{Acknowledgments}
It is a pleasure to thank Kazuya Koyama, Gustavo Niz and Ivonne Zavala for comments
on the draft. GT is supported by an STFC Advanced Fellowship ST/H005498/1.


\begin{thebibliography}{99}

\bibitem{Burgess:2013ara}
  C.~P.~Burgess,
  ``The Cosmological Constant Problem: Why it's hard to get Dark Energy from Micro-physics,''
  arXiv:1309.4133 [hep-th].

\bibitem{Clifton:2011jh}
  T.~Clifton, P.~G.~Ferreira, A.~Padilla and C.~Skordis,
  ``Modified Gravity and Cosmology,''
  Phys.\ Rept.\  {\bf 513} (2012) 1
  [arXiv:1106.2476 [astro-ph.CO]].


\bibitem{Nicolis:2008in}
  A.~Nicolis, R.~Rattazzi and E.~Trincherini,
  ``The Galileon as a local modification of gravity,''
  Phys.\ Rev.\ D {\bf 79} (2009) 064036
  [arXiv:0811.2197 [hep-th]].


\bibitem{Vainshtein:1972sx}
  A.~I.~Vainshtein,
  ``To the problem of nonvanishing gravitation mass,''
  Phys.\ Lett.\ B {\bf 39} (1972) 393.

\bibitem{deRham:2010eu}
  C.~de Rham and A.~J.~Tolley,
  ``DBI and the Galileon reunited,''
  JCAP {\bf 1005} (2010) 015
  [arXiv:1003.5917 [hep-th]].


\bibitem{Hinterbichler:2010xn}
  K.~Hinterbichler, M.~Trodden and D.~Wesley,
  ``Multi-field Galileons and higher co-dimension branes,''
  Phys.\ Rev.\ D {\bf 82} (2010) 124018
  [arXiv:1008.1305 [hep-th]].


\bibitem{VanAcoleyen:2011mj}
  K.~Van Acoleyen and J.~Van Doorsselaere,
  ``Galileons from Lovelock actions,''
  Phys.\ Rev.\ D {\bf 83} (2011) 084025
  [arXiv:1102.0487 [gr-qc]].



\bibitem{deRham:2010ik}
  C.~de Rham and G.~Gabadadze,
  ``Generalization of the Fierz-Pauli Action,''
  Phys.\ Rev.\ D {\bf 82} (2010) 044020
  [arXiv:1007.0443 [hep-th]].

\bibitem{deRham:2010kj}
  C.~de Rham, G.~Gabadadze and A.~J.~Tolley,
  ``Resummation of Massive Gravity,''
  Phys.\ Rev.\ Lett.\  {\bf 106} (2011) 231101
  [arXiv:1011.1232 [hep-th]].



\bibitem{Fubini:1976jm}
  S.~Fubini,
  ``A New Approach to Conformal Invariant Field Theories,''
  Nuovo Cim.\ A {\bf 34} (1976) 521.



\bibitem{Khoury:2011da}
  J.~Khoury, J.~-L.~Lehners and B.~A.~Ovrut,
  ``Supersymmetric Galileons,''
  Phys.\ Rev.\ D {\bf 84} (2011) 043521
  [arXiv:1103.0003 [hep-th]].

\bibitem{Koehn:2013hk}
  M.~Koehn, J.~-L.~Lehners and B.~Ovrut,
  ``Supersymmetric Galileons Have Ghosts,''
  Phys.\ Rev.\ D {\bf 88} (2013) 023528
  [arXiv:1302.0840 [hep-th]].

\bibitem{Farakos:2013fne}
  F.~Farakos, C.~Germani and A.~Kehagias,
  ``Ghost-free Supersymmetric Galileons,''
  arXiv:1306.2961 [hep-th].

\bibitem{Fairlie:2011md}
  D.~Fairlie,
  ``Comments on Galileons,''
  J.\ Phys.\ A {\bf 44} (2011) 305201
  [arXiv:1102.1594 [hep-th]].

\bibitem{Curtright:2012gx}
  T.~L.~Curtright and D.~B.~Fairlie,
  ``A Galileon Primer,''
  arXiv:1212.6972 [hep-th].

\bibitem{Deffayet:2009mn}
  C.~Deffayet, S.~Deser and G.~Esposito-Farese,
  ``Generalized Galileons: All scalar models whose curved background extensions maintain second-order field equations and stress-tensors,''
  Phys.\ Rev.\ D {\bf 80} (2009) 064015
  [arXiv:0906.1967 [gr-qc]].




\bibitem{Koyama:2011yg}
  K.~Koyama, G.~Niz and G.~Tasinato,
  ``Strong interactions and exact solutions in non-linear massive gravity,''
  Phys.\ Rev.\ D {\bf 84} (2011) 064033
  [arXiv:1104.2143 [hep-th]].



\bibitem{Padilla:2010de}
  A.~Padilla, P.~M.~Saffin and S.~-Y.~Zhou,
  ``Bi-Galileon theory I: Motivation and formulation,''
  JHEP {\bf 1012} (2010) 031
  [arXiv:1007.5424 [hep-th]].



\bibitem{Deffayet:2010zh}
  C.~Deffayet, S.~Deser and G.~Esposito-Farese,
  ``Arbitrary $p$-form Galileons,''
  Phys.\ Rev.\ D {\bf 82} (2010) 061501
  [arXiv:1007.5278 [gr-qc]].



\bibitem{Koyama:2011wx}
  K.~Koyama, G.~Niz and G.~Tasinato,
  ``The Self-Accelerating Universe with Vectors in Massive Gravity,''
  JHEP {\bf 1112} (2011) 065
  [arXiv:1110.2618 [hep-th]].


\bibitem{Gabadadze:2013ria}
  G.~Gabadadze, K.~Hinterbichler, D.~Pirtskhalava and Y.~Shang,
  ``On the Potential for General Relativity and its Geometry,''
  arXiv:1307.2245 [hep-th].

\bibitem{Ondo:2013wka}
  N.~A.~Ondo and A.~J.~Tolley,
  ``Complete Decoupling Limit of Ghost-free Massive Gravity,''
  arXiv:1307.4769 [hep-th].

\bibitem{Tasinato:2013oja}
  G.~Tasinato, K.~Koyama and N.~Khosravi,
  ``The role of vector fields in modified gravity scenarios,''
  arXiv:1307.0077 [hep-th];
  G.~Tasinato, K.~Koyama and G.~Niz,
  ``Exact Solutions in Massive Gravity,''
  Class.\ Quant.\ Grav.\  {\bf 30} (2013) 184002
  [arXiv:1304.0601 [hep-th]];
  G.~Tasinato, K.~Koyama and G.~Niz,
  ``Vector instabilities and self-acceleration in the decoupling limit of massive gravity,''
  Phys.\ Rev.\ D {\bf 87} (2013) 064029
  [arXiv:1210.3627 [hep-th]];


\bibitem{Jackiw:2011vz}
  R.~Jackiw and S.~-Y.~Pi,
  ``Tutorial on Scale and Conformal Symmetries in Diverse Dimensions,''
  J.\ Phys.\ A {\bf 44} (2011) 223001
  [arXiv:1101.4886 [math-ph]].



\bibitem{deRham:2013hsa}
  C.~de Rham, M.~Fasiello and A.~J.~Tolley,
  ``Galileon Duality,''
  arXiv:1308.2702 [hep-th].


\bibitem{Creminelli:2013fxa}
  P.~Creminelli, M.~Serone and E.~Trincherini,
  ``Non-linear Representations of the Conformal Group and Mapping of Galileons,''
  arXiv:1306.2946 [hep-th].




\bibitem{Deser:1998rj}
  S.~Deser and G.~W.~Gibbons,
  ``Born-Infeld-Einstein actions?,''
  Class.\ Quant.\ Grav.\  {\bf 15} (1998) L35
  [hep-th/9803049].


\end{thebibliography}
\end{document}